\documentclass[fleqn]{article}
\usepackage{a4wide}
\usepackage{graphicx}
\usepackage{tabularx}

\newcommand{\be}{\begin{equation}}
\newcommand{\ee}{\end{equation}}
\newcommand{\bear}{\begin{eqnarray}}
\newcommand{\eear}{\end{eqnarray}}
\begin{document}
\title{Spatiotemporal Patterns in Arrays of Coupled Nonlinear
Oscillators\thanks{Based on the Dr. Biren Roy Memorial Lecture delivered 
at Jawaharlal Nehru University by M. Lakshmanan on 27 October 1998}}
\author{M. Lakshmanan and P. Muruganandam \\Centre for Nonlinear Dynamics, 
Department of Physics\\ Bharathidasan University, Tiruchirapalli 620 024, 
India}
\date{ }
\maketitle
\begin{abstract}

Nonlinear reaction-diffusion systems admit a wide variety of
spatiotemporal patterns or structures. In this lecture, we point out
that there is certain advantage in studying discrete arrays, namely
cellular neural/nonlinear networks (CNNs), over continuous systems.
Then, to illustrate these ideas, the dynamics of diffusively coupled
one and two dimensional cellular nonlinear networks (CNNs), involving
Murali-Lakshmanan-Chua circuit as the basic element, is considered.
Pro\-pagation failure in the case of uniform diffusion and propagation
blocking in the case of defects are pointed out. The mechanism behind
these phenomena in terms of loss of stability is explained.  Various
spatiotemporal patterns arising from diffusion driven instability such
as hexagons, rhombous and rolls are considered when external forces are
absent.  Existence of penta-hepta defects and removal of them due to
external forcing is discussed. The transition from hexagonal to roll
structure and breathing oscillations in the presence of external
forcing is also demonstrated. Further spatiotemporal chaos,
synchronization and size instability in the coupled chaotic systems are
elucidated.

\end{abstract}
\tableofcontents
\section{Introduction} 
\label{sec1} 

\begin{figure}[!ht]
\begin{center}
\includegraphics[width=\linewidth]{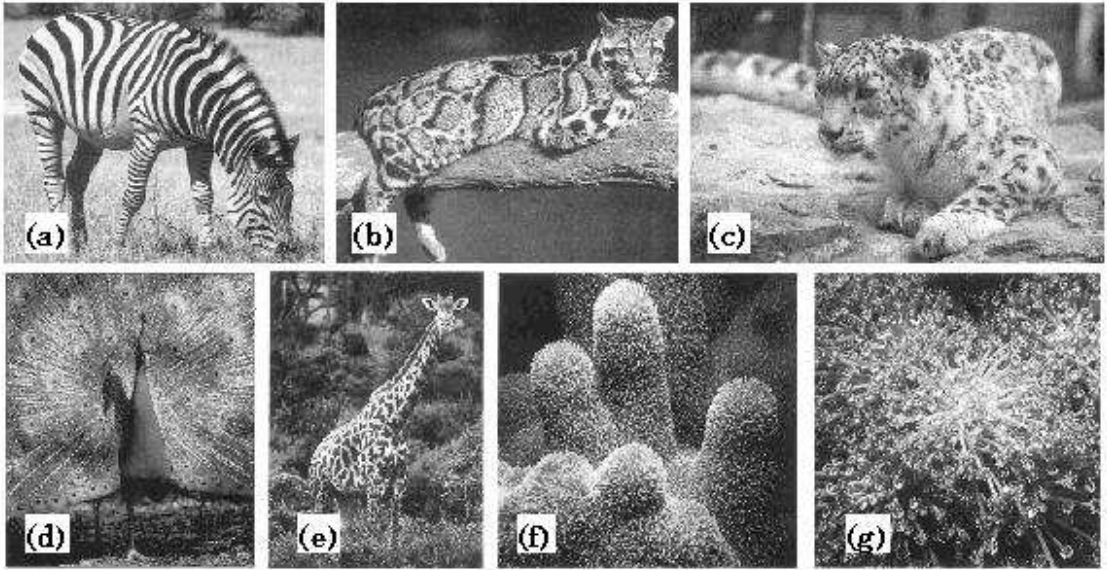}
\end{center}
\caption{A select band of patterns in natural systems: (a) Zebra (b) Panther 
(c) snow leopard (d) peacock (e) giraffe  (f) pilar coral and (e) daisy coral} 
\label{fign1} 
\end{figure}

Patterns abound in nature. If we look around us, we can observe myriads
of interesting patterns ranging from uniform to very complex
varieties.  They occur in varied phenomena encompassing physics,
chemistry, biology, social dynamics, economics and so on. Essentially
they are distinct structures on a space-time scale, which arise as a
collective and cooperative phenomena due to the underlying large number
of constituent systems. The latter could be aggregates of particles,
atoms, molecules, cells, circuits, defects, dislocations and so on.
When these aggregates can move and/or interact, they give rise to the
various patterns. A small select set of patterns is shown in
Fig.~\ref{fign1}. The patterns tell us much about the dynamics of the
macroscopic as well as to some extent the microscopic behaviours of the
underlying systems. Naturally when the interactions among the
constituents are nonlinear, one might expect novel and unexpected
patterns.

Patterns could be simple and complex. They could be stationary (eg.
still image) or changing with time (eg. recurrent image). They could
also tend towards a goal or target asymptotically. Homogeneous or
uniform patterns, though trivial, are important basic structures. One
can have often travelling wave patterns, especially in \emph{dispersive
systems}. Under suitable nonlinear forces, dispersive systems can even
admit a different kind of waves namely solitons, which are novel type
of localized spatiotemporal patterns, retaining their identity for
ever. Perturbations of them can also give rise to further interesting
structures. However, even more novel structures which can mimic
naturally occurring patterns arise when one considers nonlinear
\emph{diffusive} (especially the so called \emph{reaction-diffusion})
\emph{systems}. When large aggregates of microstructures consisting of
atoms, molecules, defects, dislocations etc are able to move and
interact, the evolution of the concentration of the species can be
shown to obey nonlinear diffusive equations of reaction type.  They may
be deduced from the underlying mass, energy, momentum, etc, balance
equations. The general form of the nonlinear reaction-diffusion
equation can be given by 
\be 
\label{eq:1} 
\frac{\partial c}{\partial t}=\vec \nabla .(D\vec \nabla c)+f(c,\vec r,t).  
\ee 
Here $c$ represents the population or concentration density of the
species and $D$ and $f$ are, in general, nonlinear functions of $c$
representing the diffusivity and the reaction kinetics, respectively.

For example, one can consider the population density of a particular
species and then $f$ in eq.~(\ref{eq:1}) will represent the birth and
death processes. In the case of logistic population growth
$f=ac(1-\delta c)$, where $a$ is the linear reproduction rate and
$\delta$ is the inhibition rate. If $D$ is a constant, then, the above
equation (\ref{eq:1}) will correspond to the celebrated Fisher
equation,
\be
\label{eq:2}
\frac{\partial c}{\partial t} = D\nabla^2 c
+ac(1-\delta c).
\ee
Originally Fisher proposed the one dimensional version of the equation 
in 1937 for the spread of an advantageous gene in a population\cite{fisher}.

\subsection{Reaction-diffusion systems}

Let us consider the situation in which two or more number of species
such as different chemicals, bacteria, populations and so on interact
and evolve. In such a situation, one may study the underlying dynamics
by considering the appropriate reaction-diffusion equations.  Examples
include the following systems.

\subsubsection{Belousov-Zhabotinsky reaction (Oregonator model)} 

This is a simple model introduced by Fields, K\"or\"os and Noyes of
University of Oregon, U.S.A. in 1972 to explain the various features of
the Belousov-Zhabotinsky reaction (for details see for
example\cite{cross}). It can be expressed in a particular form as
\bear
\label{eq:oregon}
u_{1t} & = & D_1\nabla^2u_1+\eta^{-1}\left[u_1(1-u_1)-
\frac{bu_2(u_1-a)}{u_1+a}\right],\nonumber \\
u_{2t} & = & D_2\nabla^2u_2+u_1-u_2.
\eear
Here $u_1$ represents the concentration of the autocatalytic species
$HBrO_2$ and $u_2$ is the concentration of the transition ion catalyst
in the oxidized state $Ce^{3+}$ or $Fe^{3+}$. $\eta$, $a$ and $b$ are
parameters. In eq.~(\ref{eq:oregon}) the suffix $t$ represents partial
derivative.

\subsubsection{Brusselator model} 

This is one of the often studied model for the formation of chemical
patterns introduced originally by Lefever, Nicolis and Prigogine (see
\cite{walgra1}). It is based on the following chemical reaction:
\bear
A & \longrightarrow & X, \nonumber \\
B+X & \longrightarrow & Y, \nonumber \\
2X+Y & \longrightarrow & 3X, \nonumber \\
X & \longrightarrow & E,
\eear
where the concentration of the species $A$, $B$ and $E$ are maintained
constant and are thus real control parameters of the system. After
appropriate scaling, the evolution of the active species $X$ and $Y$
can be described by the following set of equations\cite{walgra1}:
\bear
\label{eq:bruss}
\partial_t X & = & A-(B+1)X+X^2Y+D_X\nabla^2X, \nonumber\\
\partial_t Y & = & BX-X^2Y+D_Y\nabla^2Y.
\eear
Here $D_X$ and $D_Y$ are diffusion coefficients.

\subsubsection{Lotka-Volterra predator-prey model} 

A model for interacting populations of two species is one in which the
population of the prey is dependent on the predator and vice versa.
Such systems can be represented by the following set of
equations\cite{okubo}:
\bear
\label{eq:prey}
\frac{\partial S_1}{\partial t} & = & 
D_1\frac{\partial^2 S_1}{\partial x^2}
+a_1S_1-b_1S_1S_2,\nonumber \\
\frac{\partial S_2}{\partial t} & = & 
D_2\frac{\partial^2 S_2}{\partial x^2}
-a_2S_2+b_2S_1S_2.
\eear
Here $S_1$ and $S_2$ are the population densities of prey and predator,
$D_1$ and $D_2$ are the diffusivities of the two populations,
respectively.  $a_1$, $a_2$ are the linear ratio of birth and death for
the individual species; $b_1$ and $b_2$ are the linear decay and growth
factors due to interaction.

\subsubsection{Gierer-Meinhardt model for biological pattern formation}

The possible interaction of an activator and a rapidly diffusing
inhibitor can be modeled by the following set of equations\cite{koch}:
\bear
\label{eq:koch}
\frac{\partial a}{\partial t} & = & 
D_a\Delta a +\rho_a\frac{a^2}{1+\kappa_aa^2}-\mu_aa+\sigma_a,\nonumber \\
\frac{\partial h}{\partial t} & = & 
D_h\Delta h +\rho_ha^2-\mu_hh+\sigma_h.
\eear
Here $\mu_a$, $\mu_h$ are removal rates, $\rho_a$, $\rho_h$ are the
cross reaction coefficients and $\sigma_a$, $\sigma_h$ are the basic
products of the activator and inhibitor, respectively. $\kappa_a$
corresponds to the saturation constant.

\subsubsection{FitzHugh-Nagumo nerve conduction model} 

A well known model for impulse propagation along the neuronal axons of
living organisms is the FitzHugh-Nagumo model represented by the
following set of equations\cite{scott},
\bear
\label{eq:fhn}
V_t & = & V_{xx}+V-\frac{V^3}{3}-R+I(x,t),\nonumber \\
R_t & = & c(V+a-bR).
\eear
Here $V$ represents the action potential and $R$ corresponds to the
lumped refractory variable. $I(x,t)$ is the external injected current,
$a$, $b$ are positive constants and $c$ corresponds to the temperature
factor.

\section{Spatiotemporal patterns in reaction-diffusion systems}

In the above, we have mentioned a few reaction-diffusion systems that
arise under different physical, chemical and biological contexts. Next
it is of considerable interest to look into the distinct space-time
structures or spatiotemporal patterns which are admitted by these
systems. Specifically we may mention the following structures: 
\begin{itemize}
\item
uniform or homogeneous steady states (trivial) 
\item
autowaves including travelling waves 
\item
spiral waves
\item
Turing patterns (rolls, stripes, hexagons,
rhombs, etc.) 
\item
localized structures 
\item
spatiotemporal chaos
\end{itemize}
\noindent and so on.  Among these patterns some of them (eg. Turing
patterns) may also be identified as \emph{dissipative structures} as
they emerge spontaneously from homogeneous equilibrium states and
correspond to systems driven away from thermodynamic equilibrium. In
the following we will briefly discuss the various space-time
structures.

\subsection{Homogeneous patterns}

The trivial, but important, class of patterns is the homogeneous or
uniform steady states. These are the equilibrium solutions of the
governing equations. Interesting dynamical features such as the
formation of various nonhomogeneous spatiotemporal patterns will arise
when these homogeneous states lose their stability.

\subsection{Autowaves} 

Transport processes in physical, chemical and physiological systems have
been often associated with special types of waves, namely travelling
waves. In general, these types of waves are termed as \emph{autowaves}
or \emph{autonomous waves}\cite{bush}. These autowaves can propagate in an
active excitable medium at the expense of energy stored in the medium
even in the absence of external driving forces. One has to note that
dispersive systems often admit different kinds of propagating waves,
called classical waves, such as sinusoidal travelling waves, wave
packets, solitary waves, solitons and so on.  However, all these
dispersive waves carry energy and information and do not consume any
energy associated with the medium. The medium in these cases is said to
be passive. On the other hand, autowaves propagate in nonlinear
diffusive as well as dissipative systems in a self sustained manner by
inducing a local release of the stored energy and use it to trigger the
same process in the adjacent regions. The medium here is termed as
active medium.  Typical examples include the waves of combustion (eg.
flame propagating along the cord of a cracker), waves of phase
transition, concentration waves in chemical reactions, nerve impulse
propagation in neuronal axons, excitation waves in cardiac tissues,
epidemic waves in ecological populations and so on.

Few important classes of autowaves which arise in many physical,
chemical and biological reaction-diffusion systems are the travelling
wavefronts, travelling pul\-ses, travelling wave trains and so on. An
example for the travelling wavefront is the solution of the Fisher
equation (\ref{eq:2}), in one spatial dimension, of the form
\[
u=u(x-ct)\equiv u(\zeta),\;\;\;\;\zeta =x-ct.  
\] 
Here $c$ is the wave speed. Travelling pulses arise for example in the
FitzHugh-Nagumo nerve conduction model (\ref{eq:fhn}). Typical example
for the travelling wavetrain is the wave solution of $\lambda$ -
$\omega$ systems\cite{murray}.

\subsection{Spiral waves}

Another interesting class of waves which are very common in a variety
of natural systems corresponds to the \emph{spiral waves} and
\emph{scroll waves} that arise in more than one spatial dimensions.
The spiral waves which arise in two spatial dimensions can be realized
by considering a single wave propagating around a circular
obstacle\cite{avhold}. The wave repeatedly travels along the same path
at a frequency given by the wave velocity divided by the circumference
of the obstacle. If the radius is gradually decreased the frequency of
the obstacle increases until the wavefront of a new wave catches up
with the tail of the previous wave.  At this point the rate at which
the wave travels around the obstacle cannot increase further because
the new wave cannot re-excite regions that are still recovering from
the previous wave.

\begin{figure}[!ht] 
\begin{center}
\includegraphics[width=\linewidth]{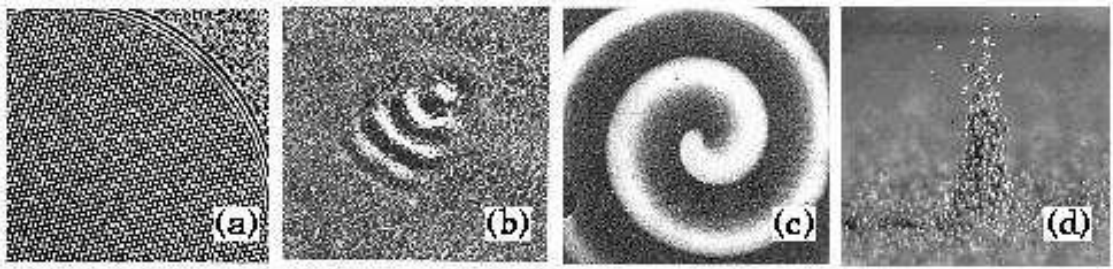}
\end{center}
\caption{(a) Hexagonal pattern in Rayleigh-B\'enard system (b) Localized 
propagating wave in binary fluid convection (c) Sprial wave in the 
Belousov-Zhabotinsky reaction and (d) Localized standing wave in vertically
vibrated layer of bronze balls} 
\label{fign2} 
\end{figure}

If the radius is made even smaller, the wave is forced to adopt a
spiral shape that continues to rotate around a central core (see Fig.
\ref{fign2}(c)). The spiral cannot enter the central core because this
region is still in the so called refractory (or recovering) state and
cannot be re-excited. An example for the spiral wave is the case of
cardiac arrhythmia in the heart during an irregularity in the
heartbeat.  The phenomenon is known to be modeled by Hodgkin-Huxley
model. Another example is the spiral patterns in the
Belousov-Zhabotinsky (BZ) reaction where scroll waves have also been
observed. Recent studies show the generation of scroll waves in
photosensitive excitable media by perturbing travelling waves to their
direction of propagation\cite{amem}

\subsection{Turing patterns}

Yet another important kind of patterns which arises in
reaction-diffusion systems is the Turing pattern. A.~M.~Turing in 1952
suggested that, under certain conditions, chemicals can react and
diffuse in such a way to produce steady state heterogeneous patterns of
chemicals\cite{turing}. He had also proposed a model for the chemical
basis of \emph{morphogenesis} (which represents the development of
structure during the growth of an organism). Turing patterns arise in
many reaction-diffusion systems when a homogeneous steady state which
is stable due to small spatial perturbations in the absence of
diffusion becomes unstable in the presence of diffusion combined with
the condition that one of the species or chemicals should diffuse
faster than the other. Typical examples include, hexagonal patterns in
Rayleigh-B\'enard system, the formation of spots in leopard, stripes in
zebra, fingerprint patterns and so on. In order to understand the
formation of Turing patterns, let us consider the model for the two
chemical species represented by the following set of rescaled
reaction-diffusion equations:
\bear 
\label{eq:tur} 
\frac{\partial
u}{\partial t} & = & \gamma f(u,v)+\nabla^2 u, \nonumber \\
\frac{\partial v}{\partial t} & = & \gamma g(u,v)+d\nabla^2 v.
\eear
Here $u$ and $v$ are the rescaled concentration of the two chemical
species, $f$ and $g$ are, in general, nonlinear functions of $u$ and
$v$ representing the kinetics. Suppose $(u_0, v_0)$ is the homogeneous
steady state, which is stable when the diffusion is absent and becomes
unstable in the presence of diffusion. Then one can write the condition
for the diffusion driven instability as\cite{murray}
\bear
\label{eq:insta} 
f_u+g_v & < & 0, \nonumber\\ 
f_ug_v-f_vg_u & > & 0,\nonumber\\
df_u+g_v & > & 0, \nonumber\\
(df_u+g_v)^2-4d(f_ug_v-f_vg_u) & > & 0, 
\eear 
where $f_u$, $f_v$, $g_u$ and $g_v$ are the derivatives of the
functions $f$ and $g$ evaluated at the steady state $(u_0,v_0)$. These
conditions can be further used for a selection of the admissible
wavenumbers associated with the instability, leading to spontaneous
formation of nontrivial spatial patterns.

\subsection{Localized structure}

Recently it has been found that certain dissipative nonlinear systems
when driven by external forces exhibit a novel class of localized
structures. Such structures have been observed in several experimental
and theoretical models\cite{umban,fine,deiss}. For example, vertically
vibrarted granuler layer shows a kind of localized oscillations called
\emph{oscillons}. These oscillons are highly localized particle-like
excitations (see Fig.~\ref{fign2}(d)) of the granular layer which
oscillate at half the driving frequency. Other examples includes
localized oscillations in fluid systems\cite{fine}, breathing solutions
of  Ginzburg-Landau equations\cite{deiss} and so on.  Essentially these
localized structures arise due to a tendency of certain nonlinear
sytems which localize dissipation in the presence of external forces.

\subsection{Spatiotemporal chaos}

The study of spatiotemporal or extensive chaos has also been receiving
considerable interest in recent times. The phenomenon of chaos in low
dimensional systems is almost well understood. However there is a lot to
be investigated in extended systems. Studies on extended systems show
that the fractal dimension increases linearly with the system size.
Further there are measures such as multiple positive Lyapunov
exponents, spatial correlation length, information functions and so on
to quantify the spatiotemporal chaos\cite{egolf,bohr}.

\section{Cellular neural/nonlinear networks (CNNs) as reaction-dif\-fusion 
systems} 

The most common factor in the autowave process and other pattern
formations in reaction-diffusion systems is the presence of active
nonlinear medium. We have already discussed the role of active medium
in the previous subsection for the autowave process. However there are
physical situations where propagation of waves is inhibited beyond
certain spatial distances in such medium, for example, the failure of
electrical impulse in the nerves of patients suffering from
\emph{multiple sclerosis}. Unfortunately there are
theorems\cite{keener} which clearly shows that continuous models cannot
exhibit such \emph{propagation failure}. In order to simulate this kind
of new phenomenon, in addition to the existing several kinds of
patterns noted above, it becomes necessary to consider discrete
versions of diffusively coupled nonlinear dynamical systems to mimic
reaction-diffusion processes. In many situations, the discrete systems
are modeled by appropriate nonlinear electronic circuits, for example
the cases of impulse propagation along nerve fiber, propagation of
action potential in cardiac tissues and so on.

Under these circumstances, it is of great interest to investigate the
dynamics of arrays of diffusively coupled nonlinear oscillators and
systems. In such cases one can often consider an array of
interconnected locally coupled cells such as neurons, nonlinear
circuits, nonlinear oscillators and so on. Such aggregates of cells may
be called \emph{cellular neural networks} (CNNs) in the case of neurons
and \emph{cellular nonlinear networks} (again CNNs) in the case of
oscillators and circuits\cite{chua1,chua2}.

\subsection{Cellular Neural Networks (CNNs)}

In general, a CNN is defined mathematically\cite{chua2} by the dynamics
of the constituent subsystems (state equations of the individual
oscillators or cells) and a \emph{synaptic law} which specifies the
interaction with their neighbours.  Such cellular neural networks,
namely, interconnections of sufficiently large number of simple
dynamical units can exhibit extremely complex, synergetic and self
organizing behaviours. Theoretically one can consider a system of
coupled ordinary differential equations (odes) to represent a
macroscopic system, in which each of the odes corresponds to the
evolution of the individual subsystem.  Such collection or aggregates
could represent the {\it discrete reaction-diffusion
systems}\cite{kura}. In other words, in reaction-diffusion systems, the
CNNs have linear synaptic law which approximates the spatial Laplacian
operator (nearest neighbour coupling). A CNN can be represented by the
following four specifications\cite{chua2}:

\begin{enumerate}
\item
The cell dynamics given by
\[
\dot x_{j} = g_j(x_j)+I_j(x_{1},x_{2} \ldots x_{j-1},x_{j},x_{j+1}\ldots 
x_{m}),\;\;\;j=1,2\ldots m, 
\]
where $I_j$ represents to the interaction between the $j^{\mbox{th}}$ 
and the remaining cells.

\item
A synaptic law representing the interaction between the cells. For 
the reaction-diffu\-sion CNN this can be given by the discrete spatial 
Laplacion operator.

\item
Appropriate boundary conditions.

\item
Initial conditions.
\end{enumerate}
Specific examples are the coupled array of anharmonic
oscillators\cite{marin}, Josephson jun\-ctions\cite{watan}, continuously
stirred tank reactors (CSTR) exhibiting travelling
waves\cite{dolnik,lapla}, propagation of nerve impulses (action
potential) along the neuronal axon\cite{scott}, the propagation of
cardiac action potential in the cardiac tissues\cite{alles} and so on.
For the past few years several investigations have been carried out to
understand the spatiotemporal behaviours of these coupled nonlinear
oscillators and systems. The studies on these systems include the
travelling wave phenomena, Turing patterns, spatiotemporal chaos and
synchronization\cite{fein,munu1,perez1,sobri,perez2,perez3,kocar}.  Of
particular interest among coupled arrays is the study of {\it
diffusively coupled driven systems} as they represent diverse topics
like Faraday instability\cite{liou}, granular
hydrodynamics\cite{umban,kudro}, self organized criticality\cite{bak}
and so on.  Identification of localized structures in these systems
has been receiving considerable attention very recently\cite{fine}.

A rather powerful and practical way of studying CNN systems is to model
the constituent cells in terms of suitable nonlinear electronic
circuits, which are then interconnected through appropriate linear
resistors. The advantages of such arrays of nonlinear electronic
circuits is that they are quite flexible, that is they can mimic real
systems but also can be studied on their own merit, they are easy to
produce and easy to study experimentally and numerically. From this
point of view already P\'erez-Mu\~nuzuri et al have studied the
dynamics of CNNs with the three variable Chua's circuit as the basic
element and identified several interesting patterns. However it will be
considerable interest if one investigates the dynamics of very simple
diffusively coupled driven nonlinear electronic circuits to realize
novel spatiotemporal patterns. Recently Murali, Lakshmanan and
Chua\cite{murali1} have introduced the simplest second order nolinear
nonautonomous disspative circuit consisting of a single nonlinear
element, namely, the Chua's diode. This simple circuit can exhibit a
variety of interesting bifurcations, chaos and so on when driven by
external periodic force\cite{murali,lak1,lak2}. Therefore it will be of
considerable interest to steady the dynamics of one and two dimensional
arrays of coupled MLC circuts. In this lecture, we wish to give brief
details of the type of spatiotemporal patterns and other features which
arise in arrays of coupled MLC circuits with and without the presence
of external force.

\subsection{Murali-Lakshmanan-Chua circuit}
\label{mlc:des}

\begin{figure}[!ht]
\begin{center}
\includegraphics[width=0.8\linewidth]{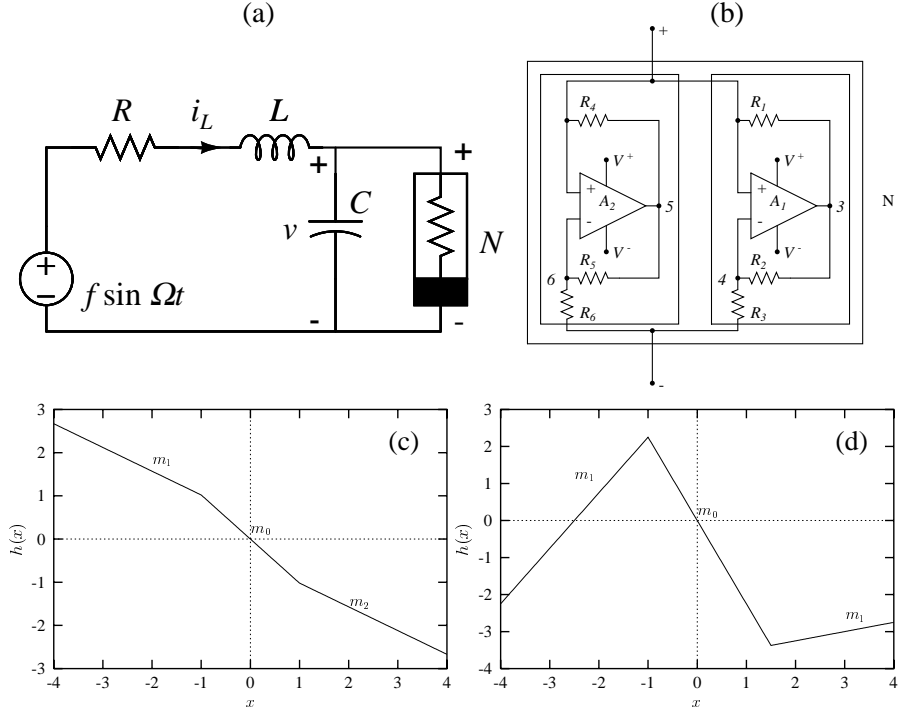}
\end{center}
\caption{(a) Circuit realization of the simple MLC circuit (b) Circuit 
realization of the Chua's diode. The characteristic
curve for the Chua's diode ($h(x)$) for (c) $\{m_0,$ $m_1,$ $m_2 \}$ $=  \{-1.02,$ 
$-.55,$ $-0.55 \}$ and (d)  $\{m_0,$ $m_1,$ $m_2 \}$ $=  \{-2.5,$ 
$1.5,$ $0.25 \}$
} 
\label{figcir} 
\end{figure}

The Murali-Lakshmanan-Chua circuit, Fig. \ref{figcir}(a), is the
simplest second order dissipative non\-autonomous circuit, consisting of
Chua's diode as the only nonlinear element\cite{murali}.
This circuit contains a capacitor $(C)$, an inductor $(L)$, a linear
resistor $(R)$, an external periodic forcing $(f\sin\Omega t)$ and a
Chua's diode.  By applying the Kirchoff's laws to this circuit, the
governing equations for the voltage $v$ across the capacitor $C$ and
the current $i_L$ through the inductor $L$ are represented by the
following set of two first order nonautonomous differential equations:
\begin{eqnarray} 
\label{app:eq1} 
C\frac{dv}{dt} & = & i_L-g(v),\nonumber \\ 
L\frac{di_L}{dt} & = & -Ri_L-R_si_L-v+f\sin\Omega t, 
\end{eqnarray} 
where $g(v)$ is a piecewise linear function
corresponding to the characteristic of the Chua's diode $(N)$ and is
given by 
\begin{equation} 
\label{app:eq2}
g(v)=\left\{
     \matrix{\epsilon'+G_2v+(G_0-G_1), & v > B_p         \cr
	     \epsilon'+G_0v,           & -B_p\leq v\leq B_p \cr
	     \epsilon'+G_1v-(G_0-G_1), & v < -B_p }
\right.
\end{equation} 

The piecewise nature of the characteristic curve of Chua's diode is
obvious from Eq. (\ref{app:eq2}). The slopes of left, middle and right
segments of the characteristic curve are $G_1$, $G_0$ and $G_2$,
respectively. $-B_p$ and $B_p$ are the break points and $\epsilon'$
corresponds to the dc offset in the Chua's diode.  Rescaling Eq.
(\ref{app:eq1}) as $v=xB_p$, $i_L=GyB_p$, $G=1/R$, $\omega=\Omega C/G$,
$t=\tau C/G$ and $\epsilon=\epsilon'/G$ and then redefining $\tau$ as
$t$ the following set of normalized equations are obtained:
\begin{eqnarray}
\label{app:eq3}
\dot x & = & y-h(x),\nonumber \\
\dot y & = & -\beta x -\sigma y +F\sin\omega t,
\end{eqnarray}
with
\begin{equation}
\label{eqn2}
h(x)=\left\{
     \matrix{\epsilon+m_2x+(m_0-m_1), & x\geq x_2         \cr
             \epsilon+m_0x,           & x_1\leq x\leq x_2 \cr
             \epsilon+m_1x-(m_0-m_1), & x\leq x_1
            }
\right., 
\end{equation}
where $\beta=(C/LG^2)$, $\sigma=(C/LG^2)(1+GR_s)$ and
$F=(f\beta/B_p)$.  Obviously $h(x)$ takes the form as in Eq.
(\ref{eqn2}) with $m_0=G_0/G$, $m_1=G_1/G$ and $m_2=G_2/G$. The
dynamics of Eq. (\ref{app:eq3}) depends on the parameters $\beta$,
$\sigma$, $m_0$, $m_1$, $m_2$, $\epsilon$, $\omega$ and $F$.

The rescaled parameters in the experimental observations correspond to
$\beta=1.0$, $\sigma=1.015$, $m_0=-1.02$, $m_1=-0.55$, $m_2-0.55$,
$\epsilon=0$ and $\omega=0.75$.  By varying $F$ one can observe the
familiar period doubling bifurcations leading to chaos and several
\begin{figure}[!ht]
\begin{center}
\includegraphics[width=0.8\linewidth]{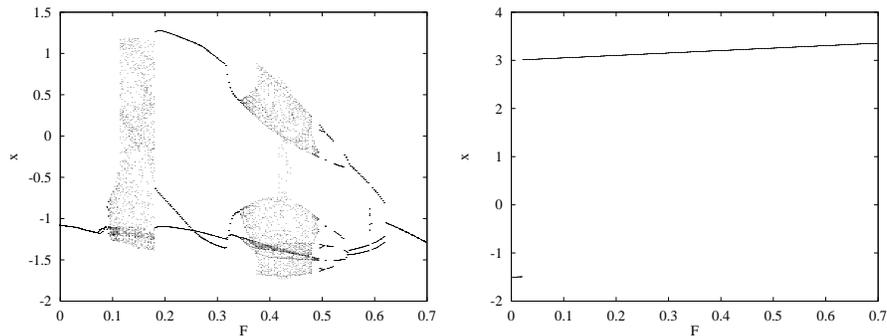}
\end{center}
\caption{ Bifurcation diagram in the $F-x$ plane (a) for $\{m_0,$ $m_1,$ $m_2,$ 
$\epsilon,$ $\beta,$ $\sigma,$ $\omega\} =  \{-1.02,$ $-.055,$ $-0.55,$ $0.0,$ 
$1.0,$ $1.015,$ $0.75 \}$ and (b) for $\{m_0,$ $m_1,$ $m_2,$ $\epsilon,$ $\beta,$ 
$\sigma,$ $\omega\} =  \{-2.25,$ $1.5,$ $0.25,$ $0.0,$ $1.0,$ $1.0,$ $0.75\}$ 
} 
\label{fig4} 
\end{figure}
periodic windows in the MLC circuit. Fig.~\ref{fig4}(a) shows the one
parameter bifurcation diagram in the $F-x$ plane $[F\in (0,0.7)]$. A
summary of bifurcations that occur in this case for different $F$
values is given in Table \ref{table1}.  Further it is of great interest
to consider the parametric choice $\{m_0, m_1, m_2, \epsilon, \beta,
\sigma, \omega\} =  \{-2.25, 1.5, 0.25, 0.0,$ $1.0, 1.0, 0.75 \}$ which
corresponds to the function $h(x)$ having the form as shown in
Fig.~\ref{figcir}(d).  This choice of parameters provides the possibility
of bistability nature in the asymmetric case in the absence of periodic
forcing. In this case one can easily observe from numerical simulations
that the MLC circuit admits only limit cycles for $F\in (0,0.7)$. The
bifurcation diagram in the $F-x$ plane is depicted in
Fig.~\ref{fig4}(b).  
\begin{table}[!ht]
\centering
\caption{Summary of bifurcation phenomena of Eq.~(13) }
\begin{tabularx}{.9\linewidth}{X|X}
\hline 
amplitude $(F)$        & description of attractor \\ 
\hline
$0     <F\leq 0.071 $  & period-1 limit cycle \\ 
$0.071 <F\leq 0.089 $  & period-2 limit cycle \\ 
$0.089 <F\leq 0.093 $  & period-4 limit cycle \\ 
$0.093 <F\leq 0.19  $  & chaos \\ 
$0.19  <F\leq 0.3425$  & period-3 window \\ 
$0.3425<F\leq 0.499 $  & chaos \\ 
$0.499 <F\leq 0.625 $  & period-3 window \\ 
$0.625 <F           $  & period-1 boundary \\
\hline
\end{tabularx}
\label{table1}
\end{table}

\subsection{Present study}

In the following we discuss the active wave propagation and various
spatiotemporal patterns associated with the one and two dimensional
CNNs represented by diffusively coupled MLC circuits. First we made a
critical review on the active wave propagation that occurs in the
autonomous system by analysing the linear stability properties of the
coupled systems.  This analysis give us a clear understanding of the
wave propagation in coupled systems which is essentially the loss of
stability of the steady states via subcritical bifurcation coupled with
the existence of necessary basin of attraction for the steady states
associated with the coupled system. We also discuss the effect of weak
coupling on the active wave propagation which causes a blocking of the
wave when it reaches the weakly coupled cell. Then we look into various
spatiotemporal patterns due to the presence of Turing instabilty in the
absence of external force.

Further as mentioned earlier it is of great physical interest to study
the dynamics of the coupled oscillators when individual oscillators are
driven by external forces.  We study the spatiotemporal patterns in the
presence of periodic external force and investigate the effect of it on
the propagation phenomenon and Turing patterns. Depending upon the
choice of control parameters, a transition from hexagons to regular
rhombic structures, hexagons to rolls and then to breathing
oscillations from hexagons are observed.  The presence of external
force with sufficient strength removes the penta-hepta defect pair
originally present in the spontaneously formed hexagonal patterns
leading to the formation of regular rhombic structures.  The inclusion
of external periodic force can also induce a transition from hexagons
to rolls provided there are domains of small roll structures in the
absence of force. We further show that in the region of Hopf-Turing
instability, the inclusion of external periodic force with sufficiently
small amplitude induces a type of breathing oscillations though the
system shows a regular hexagonal pattern in the absence of any external
force.

Finally, we also study the spatiotemporal chaotic dynamics of the one
dimensional array of MLC circuits when individual oscillators oscillate
chaotically.  In this case, the emergence of spatiotemporal patterns
depends on the system size. For larger size, above a critical number of
cells, we observe a controlled space-time regular pattern eventhough
the single MLC circuit itself oscillates chaotically. However,
synchronization occurs for a smaller system size, below the threshold
limit. 

\section{Arrays of Murali-Lakshmanan-Chua (MLC) circuits}
\label{sec2}

As mentioned in the introduction (see \ref{mlc:des}), the circuit
proposed by Murali, Lakshmanan and Chua (MLC) is one of the simplest
second order dissipative nonautonomous circuit having a single
nonlinear element\cite{murali}. Here we will consider one and two
dimensional arrays of such MLC circuits, where the intercell couplings
are effected by linear resistors.

\subsection{One-dimensional array} 
\begin{figure}[!ht]
\begin{center}
\includegraphics[width=\linewidth]{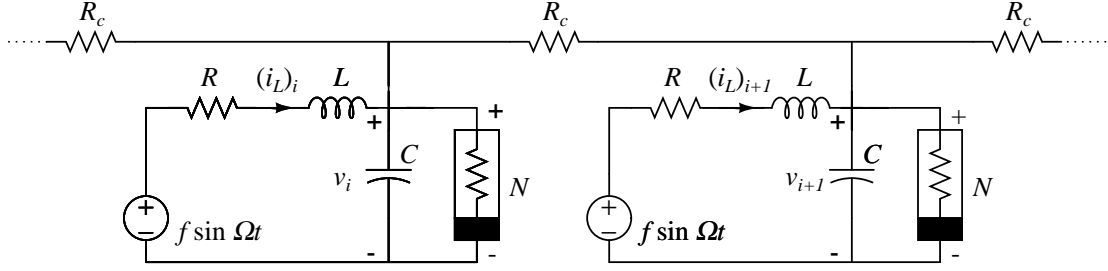}
\end{center}
\caption{Circuit diagram showing one dimesnional array of coupled MLC 
circuits} 
\label{fig2} 
\end{figure}

Fig.~\ref{fig2} shows a schematic representation of an one
dimensional  chain of resistively coupled MLC circuits.  The dynamics
of the one dimensional chain can be easily sho\-wn to be governed by the
following system of equations, in terms of suitable rescaled variables,
\begin{eqnarray}
\label{eqn1all}
\dot x_{i} & = & y_{i}-h(x_{i})+D(x_{i+1}+x_{i-1}-2x_{i}), \\
\dot y_{i} & = & -\sigma y_{i}-\beta x_{i}+F\sin \omega t, 
\,\,\,\,\, i=1,2,\cdots,N,
\end{eqnarray}
where $D$ is the diffusion coefficient, $N$ is the chain length and
$h(x)$ is a three segment piecewise linear function representing the
current voltage characteristic of the Chua's diode and given in
eq.~(\ref{eqn2}).  In (\ref{eqn2}) $m_0$, $m_1$ and $m_2$ are the three
slopes. Depending on the choice of $m_0$, $m_1$ and $m_2$ one can fix
the characteristic curve of the Chua's diode.  Here $\epsilon $
corresponds to the dc offset. In our analysis, we will consider a few
typical forms of $h(x)$ for which the coupld MLC circuits exhibit
interesting dynamics such as active wave propagation, Turing patterns,
spatiotemporal chaos and so on. In the following studies we will
consider a chain of $N=100$ MLC circuits.

\subsection{Two-dimensional array} 

As in the case of the one dimensional array introduced above, one can
also consider a two dimensional array with each cell in the array being
coupled to four of its nearest neighbours with linear resistors. The
model equation can be now written in dimensionless form as
\begin{eqnarray}
\label{eqn3all}
\dot x_{i,j} & = & y_{i,j}-h(x_{i,j})+D_1(x_{i+1,j}+x_{i-1,j}+x_{i,j+1}
+x_{i,j-1}-4x_{i,j}) \nonumber \\
& \equiv & f(x_{i,j},y_{i,j}) , \\
\dot y_{i,j} & = & -\sigma
y_{i,j}-\beta x_{i,j}+D_2 (y_{i+1,j}+y_{i-1,j}+y_{i,j+1}
+y_{i,j-1}-4y_{i,j})+F\sin \omega t\nonumber \\
& \equiv  & g(x_{i,j},y_{i,j}),\\
i,j & = & 1,2,\cdots, N.\nonumber 
\end{eqnarray}
\noindent This two dimensional array has $N\times N$ cells arranged in a
square lattice.  In our numerical study we will again take $N=100$.

In the following sections we present some of the interesting dynamical
features exhibited by the above arrays of coupled MLC circuits such as
active wave propagation and its failure, effect of weak coupling in the
propagation, Turing patterns, effect of external periodic forcing on
the Turing patterns and spatiotemporal chaotic dynamics. We have used
\emph{zero flux} boundary conditions for the study of propagation
phenomenon and Turing patterns and \emph{periodic} boundary conditions
for the study of spatiotemporal chaos in our analysis.

\section{Spatiotemporal patterns in autonomous CNNs: wave phenomena and 
Turing patterns}
\label{sec3}

Transport processes in living tissues, chemical and physical systems
have been often found to be associated with a special type of waves,
namely, active waves. In the previous section we have seen a class of
such active waves which arise in many reaction-diffusion systems. The
important aspect in the reaction-diffusion CNNs is the wave propagation
failure which occurs when the interconnections are weak, that is, for
low values of coupling strength, a feature which \emph{can not} be
realized in the continuum limit as proved by Keener\cite{keener} in
continuous homogeneous reaction-diffusion systems.  Interestingly,
experiments on biological systems shows that there exists a variety of
situations where such failure in wave propagation do arise. We will
identify such possibilities in the arrays of coupled MLC circuits.

\subsection{Active wave propagation and its failure in one dimensional
CNNs}
\label{sec3:1}

To illustrate the wave propagation failure we consider the CNN model
described by the array of coupld MLC circuits in one dimension,
Eqs.~(\ref{eqn1all}). For this purpose we have numerically integrated
Eqs.~(\ref{eqn1all}) using fourth order Runge-Kutta method with fixed
step.  In this analysis we fix the parameters at $\{\beta, \sigma, m_0,
m_1, m_2, \epsilon,$ $F \}= $ $\{1.0,$ $1.0,-2.25, 1.5, .25, 0,$ $0\}$
so that the system admits bistability which is a necessary condition to
observe a wave front. Zero flux boundary conditions are used in the
numerical computations, which in this context mean setting $x_0=x_1$
and $x_{N+1}=x_N$ at each integration step; similar choice has been
made for the variable $y$ also. To start with, we will study in this
section the autonomous case ($F=0$) and extend our studies to the
nonautonomous case $(F\neq 0)$ in the next section.

The choice of the values of the parameters guarantees the existence of
two stable equilibrium points $P^+_i =\{\sigma (m_1 - m_0 - \epsilon)/
(\beta + m_2 \sigma )$, $\beta (m_0 - m_1 - \epsilon)/ (\beta + m_2
\sigma )\}$ and $P^-_i = \{\sigma (m_1 - m_0 - \epsilon)/(\beta +
m_1\sigma )$, $ \beta (m_0 - m_1 - \epsilon)/(\beta + m_1\sigma )\}$
for each cell, $i=1,2,\ldots, 100$.  In the particular case
corresponding to  the above parametric choice, each cell in the array
\begin{figure}[!ht]
\includegraphics[width=\linewidth]{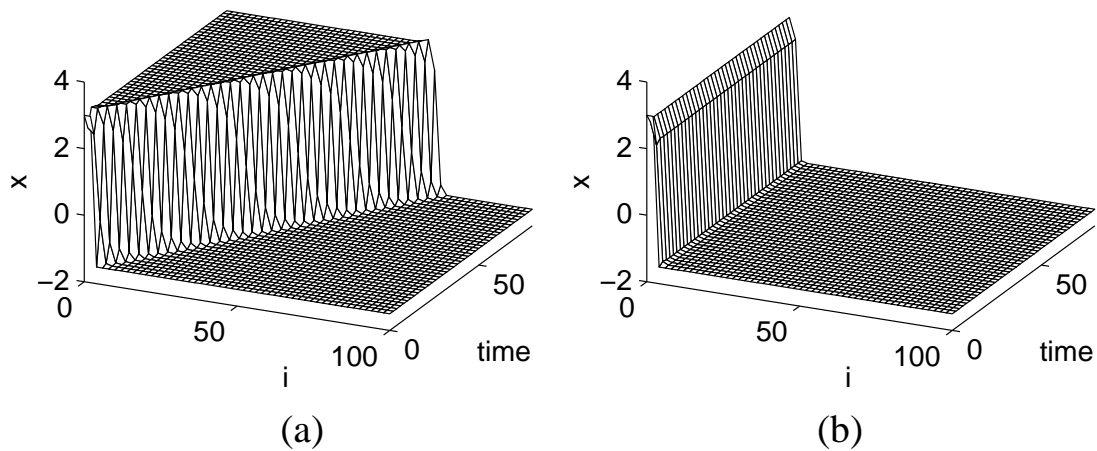}
\caption{
Space-time plot showing (a) propagation of wave fronts in one
dimensional array (100 cells) of MLC circuits for $D=2.0$ and (b)
propagation failure for $D=0.4$.
}
\label{fig:prop}
\end{figure}
has three equilibrium points $P^+_i=(3.0,-3.0)$, $P^-_i=(-1.5,1.5)$ and
$P^0_i=(0,0)$. Out of these three equilibrium points, $P^+_i$ and
$P^-_i$ are stable while $P^0_i$ is unstable.  Due to the asymmetry in
the function $h(x)$ for the present parametric choices, the basin of
attraction of the point $P^+_i$ is much larger than that of $P^-_i$ and
it is harder to steer a trajectory back into the basin $P^-_i$ once it
is in the basin of $P^+_i$.

Now we choose an initial condition such that the first few cells in the
array are excited to the $P_i^+$ state (having a large basin of
attraction compared to that of $P_i^-$) and the rest are set to $P_i^-$
state. In other words a wave front in the array is initiated by means
of the two stable steady states. On actual numerical integration of
Eqs.(\ref{eqn1all}) with $N=100$ and with the diffusion coefficient
chosen at a higher value, $D=2.0$, a motion of the wave front towards
right (see Fig.~\ref{fig:prop}(a)) is observed, that is a travelling
wave front is found.  After about 80 time units the wave front reaches
the $100^{\mbox{th}}$ cell so that all the cells are now settled at the
more stable state $(P_i^+)$ as demonstrated in Fig.~\ref{fig:prop}(a).
When the value of $D$ is decreased in steps and the analysis is
repeated, the phenomenon of travelling wavefronts continues to be
present.

However, below a critical value of the diffusion coefficient $(D=D_c)$
a failure in the propagation has been observed, which in the present
case turns out to be $D=D_c=0.4$. This means that the initiated
wavefront is unable to move as time progresses and Fig.
\ref{fig:prop}(b) shows the propagation failure for $D=0.4$.

\subsection{Propagation failure mechanism: A case study}
\label{sec3:2}

In the above, we have discussed the phenomenon of wavefront propagation
and its failure in the one dimensional array of MLC circuits by
numerically integrating eqs.~(\ref{eqn1all}) for $N=100$ cells. In
order to understand the mechanism behind wave propagation for large
coupling strengths and its failure for low values of the intercell
coupling, it is more advantageous to deal with fewer number of cells
such as $N=3,4,5,6$,etc. rather than $N=100$ because of the
diffuculties in handling large number of cells and equations
analytically. For example one has to deal with the existence of $3^N$
steady states, which is quite large for $N=100$, while it is manageble
for $N=3,4$ or $5$. With this fact in mind, we analyse analytically the
case $N=3$ and investigate the stability of the stationary states in
order to understand the nature of bifurcations. A more detailed
calculation for $N=5$ is given elsewhere\cite{muru}. One can easily
check that for $N=3$ there are $3^3=27$ steady states. They may be
obtained from the following set of defining equations:
\bear
\label{eqsta}
\dot x_1 & = & y_1-h(x_1)+D(x_2-x_1), \nonumber \\
\dot x_2 & = & y_2-h(x_2)+D(x_1-2x_2+x_3), \nonumber \\
\dot x_3 & = & y_3-h(x_3)+D(x_2-x_3), \nonumber \\
\dot y_j & = & -\sigma y_j-\beta x_j,\,\,\,\,\,\,j=1,2,3,
\eear
where $h(x)$ is as given in Eq.  (\ref{eqn2}) and the parameters are
fixed at $\beta=1$, $\sigma=1$, $m_0=-2.25$, $m_1=1.5$, $m_2=0.25$ and
$\epsilon=0$.

\subsubsection{Linear stability analysis}
\label{sec3:2:1}

Now let us consider the wave propagation. It is well known that the
propagation failure is due to the existence of a large number of
stationary states in the system\cite{keener}. Since we have considered
the coupled system with three oscillators only there exists relatively
few steady states. As we are looking for the wave front solution, it is
sufficient to analyse the stability of only a subset of the possible
$27$ states. This can be done by considering six possible steady states
in the following way: (i) the ultimate steady states, namely, $X_P^+$,
$X_P^0$ and $X_P^-$, which exist even in the absence of coupling and
(ii) the next three steady states which form wave front solutions,
namely, $X_S^+$, $X_S^0$ and $X_S^-$. The steady states $X_S^+$,
$X_S^0$ and $X_S^-$ can be found by assuming the first cell in the
$P^+$ state and the last cell in the $P^-$ state and allowing the
middle cell to be in one of the three available states, $P^+$, $P^0$,
$P^-$.  For example, $X_S^+$ state is obtained by considering the
middle cell in the $P^+$ state. In a similar fashion the rest of the
states, $X_S^0$ and $X_S^-$, are obtained. All the six steady states
can be explicitly given as below:
\bear
X_P^+ & = & \{3,3,3,-3,-3,-3\}\nonumber \\ 
X_P^0 & = & \{0,0,0,0,0,0\} \nonumber \\ 
X_P^- & = &\{-1.5,-1.5,-1.5,1.5,1.5,1.5\} 
\eear 
\bear X_S^+ & = &
\{x_{1s}^+,x_{2s}^+,x_{3s}^+,y_{1s}^+,y_{2s}^+,y_{3s}^+\}\nonumber \\
X_S^0 & = & \{x_{1s}^0,x_{2s}^0,x_{3s}^0,y_{1s}^0,y_{2s}^0,y_{3s}^0\}\nonumber \\
X_S^- & = & \{x_{1s}^-,x_{2s}^-,x_{3s}^-,y_{1s}^-,y_{2s}^-,y_{3s}^-\}
\eear
 where 
\bear 
&x_{1s}^+  = \displaystyle \frac{3(8D^2+70D+25)}{32D^2+70D+25},\nonumber  &
x_{2s}^+  =   \displaystyle\frac{3(8D^2+40D+25)}{32D^2+70D+25},\nonumber \\ 
&x_{3s}^+  =  \displaystyle\frac{1.5(16D^2-40D-25)}{32D^2+70D+25},  & 
y_{is}^+  = -x_{is}^+, \;\;\; i=1,2,3,\nonumber \\ 
&x_{1s}^0  = \displaystyle \frac{15(6D-5)}{16D^2+10D-25},\nonumber  &
x_{2s}^0  =   \displaystyle\frac{30D}{16D^2+10D-25},\nonumber \\ 
&x_{3s}^0  =  \displaystyle\frac{-7.5(4D-5)}{16D^2+10D-25},  & 
y_{is}^0  = -x_{is}^0, \;\;\; i=1,2,3,\nonumber \\ 
& x_{1s}^-  = \displaystyle -\frac{0.6(4D^2-20D-25)}{4D^2+10D+5}, &
x_{2s}^-  =  \displaystyle -\frac{0.3(8D^2+20D+25)}{4D^2+10D+5},  \nonumber \\ 
& x_{3s}^-  =  \displaystyle -\frac{0.3(8D^2+50D+25)}{4D^2+10D+5}, &
y_{is}^-  =-x_{is}^-, \;\;\; i=1,2,3.
\eear 
The stability of the above steady states can be found by linearising
(\ref{eqsta}) about them, which leads to the following
Jacobian matrix:
\be \label{jacob} J = \left (
\begin{array}{rrrrrr} 
-D-\frac{\partial h}{\partial x_1} & D & 0 & 1 &0 & 0 \\ 
D & -2D-\frac{\partial h}{\partial x_2} & D & 0 & 1 & 0 \\ 
0 &D & -D-\frac{\partial h}{\partial x_3} & 0 & 0 & 1 \\ 
-1 & 0 & 0 & -1 &0 & 0 \\ 
0 & -1 & 0 & 0 & -1 & 0 \\ 
0 & 0 &-1 & 0 & 0 & -1 
\end{array}
\right ), 
\ee 
where the derivatives, $\frac{\partial h}{\partial x_1}$,
$\frac{\partial h}{\partial x_2}$ and $\frac{\partial h}{\partial x_3}$
are evaluated at the steady states. The stability is guaranteed if all
the eigenvalues of the Jacobian matrix (\ref{jacob}) have negative real
parts. One can easily find that the states  $X_P^+$ and $X_P^-$ are
asymptotically stable and $X_P^0$ is unstable irrespective of the
values of $D$. However, on a careful analysis on the eigenvalues of the
Jacobian for the nontrivial steady states, one can find the following:
the state $X_S^+$ is asymptotically stable for $D<0.3917$ and $X_S^-$
is asymptotically stable for $D<0.5178$ while $X_S^0$ is unstable
irrespective of the $D$ values. Thus any initial condition in the
neighbourhood of $X_s^+$ ($X_S^-$) will eventually return to $X_S^+$
($X_S^-$) as long as $0 < D \le 0.3916$ ($0 < D \le 0.5177$). Further,
an arbitrary initial condition will evolve to a steady state which lies
in the neighbourhood of it. As a consequence, the system with a low
value of the diffusion coefficient for an initial condition lying in
the basin of attraction of $X_S^+$ or $X_S^-$ will transit to the
corresponding stable states $X_S^+$ or $X_S^-$ respectively and so a
failure of propagation occurs. However, for $D>D_c$, the relevant
steady loses stability and consequently the systems transits to a
suitable final steady state which is stable and this turns out to be
$X_P^+$. As a result wavefront propagation occurs. Further it may so
happen that a given initial condition may or may not lie in the basin of
attraction of a particular steady state $X_S^+$ or $X_S^-$, for a
specific value of $D$ in which case propagation does occur. Taking this
fact also into account, one can prove that for the chosen initial
conditions described in the numerical analysis above one finally
obtains $D=D_c=0.4$. Thus the wave front propagation occurs when the
steay states lose their stability via a subcritical bifurcation.

The above analysis has been extended to the cases $N=4,5$ or $6$ and
one indeed draws the same conclusion as above. Thus we can explain the
propagation failure at a critical value $D=D_c$ as due to the existence
of stationary states combined with their basin of attraction.

\subsection{Effect of weak coupling}
\label{sec3d}

In the above, the investigation has been made by considering the system
as an ideal one (as far as the circuit parameters are concerned).  But
from a practical point of view, there are defects in the coupling
parameters which may result in a weak coupling  at any of the cell in
the array. In the following we study the effect of such weak coupling
on the propagation of wave front.

Let us consider a weak coupling at the $k^{\mbox{th}}$ cell. By this we
mean that the $k^{\mbox{th}}$ cell in the array is coupled to its
nearest neighbour $(k+1)^{\mbox{th}}$ and $(k-1)^{\mbox{th}}$ cells by
resistors with slightly higher values than that of the others.  We have
studied the effect of this defect on the propagation of wave fronts.
Numerical simulations have been carried out by considering an one
dimensional array with 100 cells, where the initial conditions are
chosen as in the case of propagation phenomenon in regular one
dimensional arrays (see Sec. \ref{sec3:1}).  From the numerical
simulation results, we find that there is an abrupt stop in the
propagation when the wave front reaches the weakly coupled cell. This
happens when the coupling coefficient on either side of the
$k^{\mbox{th}}$ cell in the array has a value even above the critical
\begin{figure}[!ht]
\includegraphics[width=\linewidth]{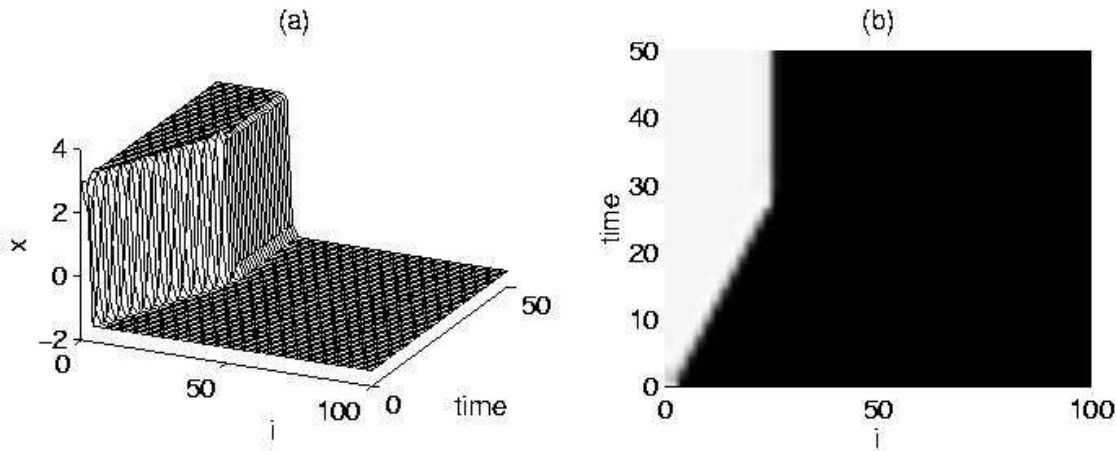}
\caption{
Figure showing the effect of weak coupling at the 25$^{\mbox{th}}$ cell
in the one dimensional array (Eq. (16)). (a) three dimensional
space-time plot and  (b) density plot.
}
\label{fig:weak}
\end{figure}
value $(D=0.4)$ for propagation failure discussed in Sec.
\ref{sec3:1}.  Fig. \ref{fig:weak} shows a blocking in the propagation
of the wave front when the coupling coefficients on either side of the
$k^{\mbox{th}}$ cell $(k=25)$, which we call as $D_k$, is set to
$0.47173$ with the rest of the coupling coefficients set to $1$
$(D=1)$.  We observe that the actual blocking occurs when the wave
front reaches the $k^{\mbox{th}}$ cell. We can say that this is a kind
of failure in the propagation because the wavefront will never reach
the last cell (that is the $100^{\mbox{th}}$ cell in the array) by
means of blocking.  Here the existence of stable steady states at the
$k^{\mbox{th}}$ cell due to the weak coupling is responsible for the
blocking of wave front. A stability analysis in the present case can be
done as in the defect free case and one can conclude that the
subcritical bifurcation is the cause of blocking.

\subsection{Turing patterns}
\label{sec3:3}

Another interesting dynamical phenomenon in the coupled arrays is the
formation of Turing patterns. These patterns are observed in many
reaction diffusion systems when a homogeneous steady state which is
stable due to small spatial perturbations in the absence of diffusion
becomes unstable in the presence of diffusion\cite{turing}.  To be
specific, the Turing patterns can be observed in a two variable
reaction-diffusion system when one of the variables diffuses faster
than the other and undergo Turing bifurcation, that is, diffusion
driven instability\cite{murray,turing}.

Treating the coupled array of MLC circuits as a discrete version of a
reaction-diffusion system, one can as well observe the Turing patterns
in this model also.  For this purpose, one has to study the linear
stability of system (3) near the steady state.  In continuous systems,
the linear stability analysis is necessary to arrive at the conditions
for diffusion driven instability. A detailed derivation of the general
conditions for the diffusion driven instability can be found in
Murray\cite{murray}. For discrete cases one can follow the same
derivation as in the case of continuous systems by considering
solutions of the form $\exp i(kj-\lambda t)$\cite{murray,munu2}. Here
$k$ and $\lambda$ are considered to be independent of the position $j$
$(j=1,2,\cdots,N)$.  For Eq.  (3), the criteria for the diffusion
driven instability can be derived by finding the conditions for which
the steady states in the absence of diffusive coupling are linearly
stable and become unstable when the coupling is present.  One can
easily show the eigenvalues that guarantee the linear stability in the
absence of coupling are the roots of the characteristic equation
\begin{equation}
\lambda^2-(f_x+g_y)\lambda+f_xg_y-f_yg_x=0, 
\end{equation}
where $f_x$, $f_y$, $g_x$ and $g_y$ are the partial derivatives of $f$
and $g$ in Eq. (3) without coupling coefficients ($D_1=D_2=0$) and
evaluated at the steady state. It can be further seen easily that the
steady state is stable in the absence of coupling if and only if the
roots of (4) ($\lambda_1$ and $\lambda_2$) have negative real parts.

Apart from the above condition, in order to satisfy the instability in
the presence of coupling (Turing instability), atleast one of the roots
of the characteristic equation,
\begin{equation}
\lambda_s^2-[k^2(D_1+D_2)-(f_x+g_y)]\lambda_s+m(k^2)=0,
\end{equation}
with
\[
m(k^2)=D_1D_2k^4-(D_2f_x+D_1g_y)k^2+f_xg_y-f_yg_x,
\]
should have positive real part.

A straightforward calculation shows that the following conditions should be
satisfied for the general reaction-diffusion system of the form given
by Eq. (\ref{eqn3all}):
\begin{eqnarray}
\label{cond:1}
f_x+g_y & < & 0,\nonumber \\
f_xg_y-f_yg_x & > &  0,\nonumber \\
f_xD_2-g_yD_1 & > & 0,\nonumber \\
(f_xD_2-g_yD_1)^2-4D_1D_2(f_xD_2-g_yD_1) & > & 0.
\end{eqnarray}
The critical wave number for the discrete system (3) can be obtained as
\begin{equation}
\label{cond:2}
\cos(k_c)=1-\displaystyle{f_xD_2-g_yD_1 \over 4D_1D_2}.
\end{equation}
Combining Eqs. (\ref{cond:1})-(\ref{cond:2}), one obtains\cite{muru}
the condition for the Turing instability such that
\begin{equation}
\label{cond:3}
\displaystyle{f_xD_2-g_yD_1 \over 8D_1D_2} \le 1.
\end{equation}

\begin{figure}[!ht]
\includegraphics[width=\linewidth]{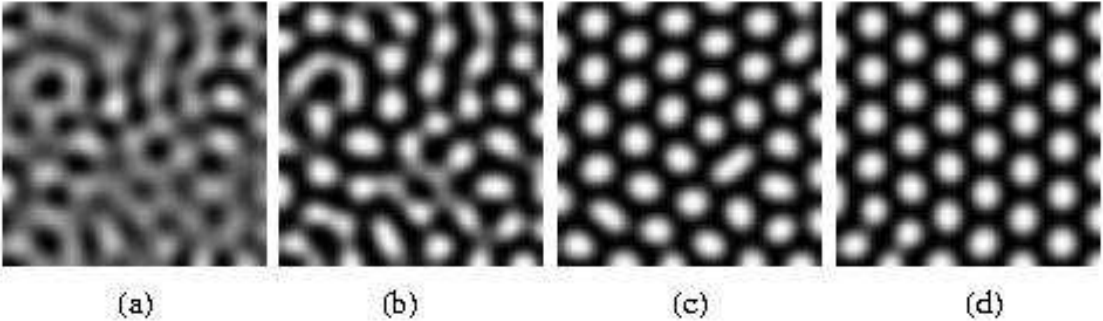}
\caption{
The spontaneous formation of Turing patterns in an array of $100\times 100$
oscillators for the parameters $\beta=0.8$, $\sigma=1$, $m_0=-0.5$, 
$m_1=0.5$, $m_2=0.5$, $\epsilon=0.1$, $F=0.0$, $\omega=0.75$, $D_1=1$ and 
$D_2=10$ in Eq. (3) at various time units (a) $T=50$, (b) $T=100$, (c) 
$T=500$ and (d) $T=2000$ (e) the same figure (d) in the Fourier space.
}
\label{fig5}
\end{figure}

We have applied these conditions to the coupled oscillator system of
the present study. For this purpose, we have fixed the parameters for
the two-dimensional model (3) as $\{\beta, \sigma,\epsilon, m_0,$ $m_1,
m_2\}=\{0.8,$ $ 0.92, 0.1, -0.5, 0.5, 0.5\}$ and verified that this
choice satisfies the conditions (\ref{cond:1}) to (\ref{cond:3}).  The
numerical simulations have been carried out using an array of size
$100\times 100$ and random initial conditions near the steady states
have been chosen for the $x$ and $y$ variables.  Figs.\
\ref{fig5}(a)-\ref{fig5}(d) show how the diffusion driven instability
leads to stable hexagonal pattern (Fig. \ref{fig5}(d)) after passing
through intermediate stages (Figs.  \ref{fig5}(a)-\ref{fig5}(c)).
Further, the spontaneously formed patterns are fairly uniform hexagonal
patterns having a penta-hepta defect pair. These defects are inherent
and very stable. 

\section{Spatiotemporal patterns in the presence of periodic external
force} 
\label{sec4}

The effect of external fields on a variety of dynamical systems has
been studied for a long time as driven systems are very common from a
practical point of view. These systems can be either spatially
modulated or temporally modulated. For example, in a large number of
dynamical systems including the Duffing oscillator, van der Pol
oscillator and the presently studied MLC circuits, temporal forcing
leads to a variety of complex dynamical phenomena including
bifurcations and chaos. In particular, it has been shown that the
resonant coupling between the forcing and the oscillatory modes may
lead to several complex dynamical patterns including quasiperiodicity,
intermittency and chaos\cite{lak2,guck,hao}.  Also the studies on the
effect of external fields in spatially extended systems have been
receiving considerable interest in recent times\cite{walgra1,walgra2,pismen}. 
Particularly, with the recent advances in identifying localized and
oscillating structures and other spatiotemporal patterns in driven
nonlinear dissipative systems such as granular media, driven
Ginzburg-Landau equations and so on, it is of special interest to study
the effects of forcing on arrays of coupled systems such as (3).
Motivated by the above, we investigate the effect of external forcing
on the propagation of wave front and formation of Turing patterns in
the coupled MLC circuits in one and two dimensions.

\subsection{Effect of external forcing on the propagation of
wave fronts}
\label{sec4:1}

\begin{figure}[!ht]
\centering
\includegraphics[width=\linewidth]{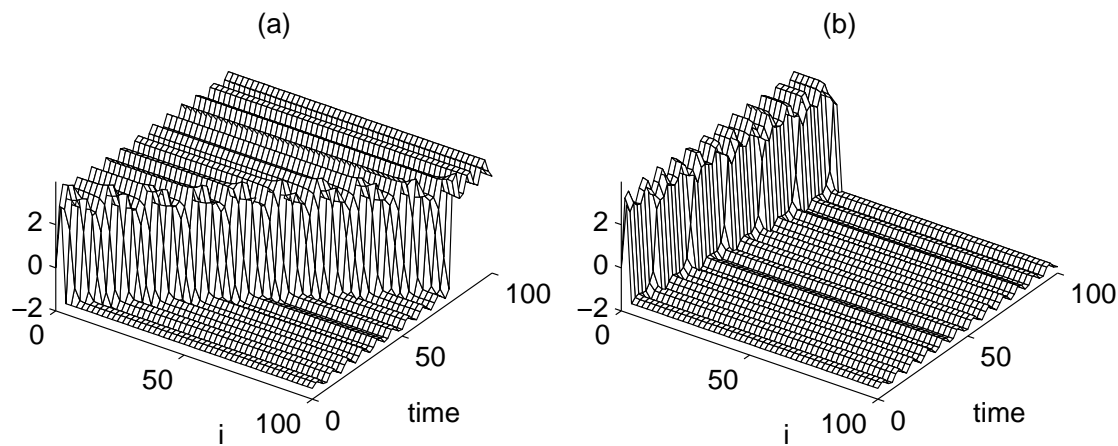}
\caption{
Propagation phenomenon in the presence of forcing: (a) Propagation
of wavefront for $D=2.0$ and $F=0.6$ and (b) The partial propagation
observed for $D=0.22$ and $F=0.6$.
}
\label{fig6}
\end{figure}

In this subsection we study the effect of external forcing on the
propagation of wave fronts. For this purpose we consider an one
dimensional array of coupled MLC circuits with initial and boundary
conditions as discussed in Sec. \ref{sec3:1}.  Now we perform the
numerical integration by the inclusion of external periodic force of
frequency $\omega=0.75$ in each cell of the array (see Eq. (1)).  By
varying the strength, $F$, of the external force we study the behaviour
of the propagating wave front in comparison with the force free case
($F=0$) as discussed in Sec. \ref{sec3:1}. We find that in the
propagation region ($D>D_c$), the effect of forcing is just to
introduce temporal oscillations and the propagation continues without
any disturbance (see Fig. \ref{fig6}(a)) as in the case $F=0$ (Fig.
\ref{fig:prop}(a)).  Of course this can be expected as the external force
is periodic in time.  However, interesting things happen in the
propagation failure region discussed in Sec. \ref{sec3:1}. In this
region, beyond a certain critical strength of the external forcing, the
wavefront tries to move a little distance and then stops, leading to a
partial propagation. Fig. \ref{fig6}(b) shows such a partial
propagation observed for $F=0.6$ and $D=0.22$ (This may be compared
with Fig. \ref{fig:prop}(b)). The phenomenon can be explained by
considering the propagation failure mechanism discussed in Sec.
\ref{sec3:2} in which one may look for a spatially stationary and
temporally oscillating wavefront. The initial wavefront tries to settle
in the nearby stationary state. However, the system will take a little
time and space to settle due the effect of forcing combined with the
transient behaviour of the system. Thus the inclusion of external
forcing in the propagation failure region can induce the wavefront to
achieve a partial propagation.

\subsection{Transition from hexagons to rhombs}
\label{sec4:2}

It is well known that the defects are inherent in very many natural
pattern forming systems. In most of the pattern forming systems, the
observed patterns are not ideal. For example, the patterns are not of
perfect rolls or hexagons or rhombs. A commonly observed defect in such
systems is the so called penta-hepta defect (PHD) pair which is the
bound state of two dislocations\cite{tsim}. Experiments on spatially
extended systems often show the occurrence of PHD in spontaneously
developed hexagonal patterns\cite{panta} In the present case also, the
existence of PHD pair can be clearly seen from Fig. \ref{fig5}(d).  In
such a situation, it is important to study the effect of external
periodic force in the coupled arrays of MLC circuits.
\begin{figure}[!ht]
\centering
\includegraphics[width=0.25\linewidth]{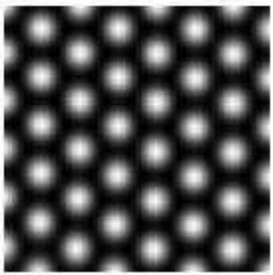}
\caption{
Figure showing the perfect organization of rhombic array in the presence 
of external periodic force with $F=0.25$.
}
\label{fig7}
\end{figure}

Now we include a periodic force with frequency $\omega$ and amplitude
$F$ in each cell of the array and we numerically integrate
Eqs.(\ref{eqn3all}) using fourth order Runge-Kutta method with zero
flux at boundaries. By fixing the frequency of the the external
periodic force as $\omega=0.75$ and varying the amplitude ($F$) we
analyse the pattern which emerges spontaneously.  Interestingly for
$F=0.25$, the defects (PHD pair) which are present in the absence of
external force (Fig. \ref{fig5}(d)), gets removed resulting in the
transition to a regular rhombic array.  Fig. \ref{fig7} shows the gray
scale plot of the pattern observed for $F=0.25$.  Thus, from the above
we infer that the inclusion of external periodic force can cause a
transition from hexagonal pattern to rhombic structures.

\subsection{Transition from hexagons to rolls}
\label{sec4:3}

\begin{figure}[!ht]
\centering
\includegraphics[width=\linewidth]{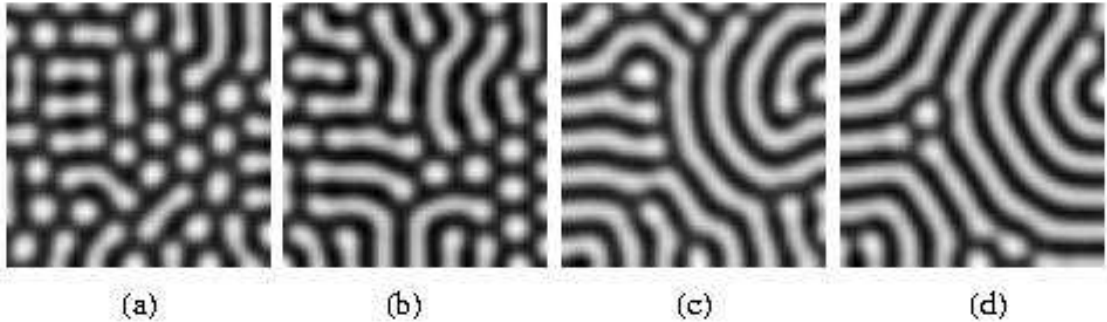}
\caption{
Figure showing the transition from hexagons to rolls for  $\{ \beta,$ 
$\sigma,$ $\epsilon,$ $m_0$, $m_1,$ $m_2\}$ $=$ $\{0.734722,$ $0.734722,$ 
$0.15,$ $-0.874,$ $-0.4715,$ $-0.4715\}$ with $D_1=1$ and $D_2=5$. (a) 
$F=0$, (b) $F=0.15$, (c) $F=0.35$ and (d) $F=0.45$.
}
\label{fig8}
\end{figure}

In addition to the transition from hexagons to rhombs by the influence
of external periodic force, there are also other possible effects due
to it. To realize them, we consider a different set of parametric
choice $\{\beta, \sigma, m_0, m_1, m_2\}$ $=$ $\{0.734722,$
$0.734722,$  $-0.874,$ $ -0.4715,$ $-0.4715\}$ with $\epsilon=0.15$,
$D_1=1.0$ and $D_2=5.0$.  For this choice the system shows hexagonal
patterns with defects including domains of small roll structures (Fig.
\ref{fig8}(a)).

Now when the external periodic force is included a transition in the
pattern from hexagonal structure to rolls starts appearing.  By fixing
the frequency of the external periodic force again at $\omega=0.75$, we
observed the actual transition from hexagons to rolls as we increase
the forcing amplitude ($F$).  Figs. \ref{fig8}(b)-\ref{fig8}(d) show
the gray scale plots for $F=0.15$, $0.35$ and $0.45$, respectively.
Obviously the transition is due to the existence of small roll
structures in the pattern for $F=0$ which nucleates the formation of
rolls in the presence of forcing.

\subsection{Breathing oscillations}
\label{sec4:4}

\begin{figure}[!ht]
\centering
\includegraphics[width=\linewidth]{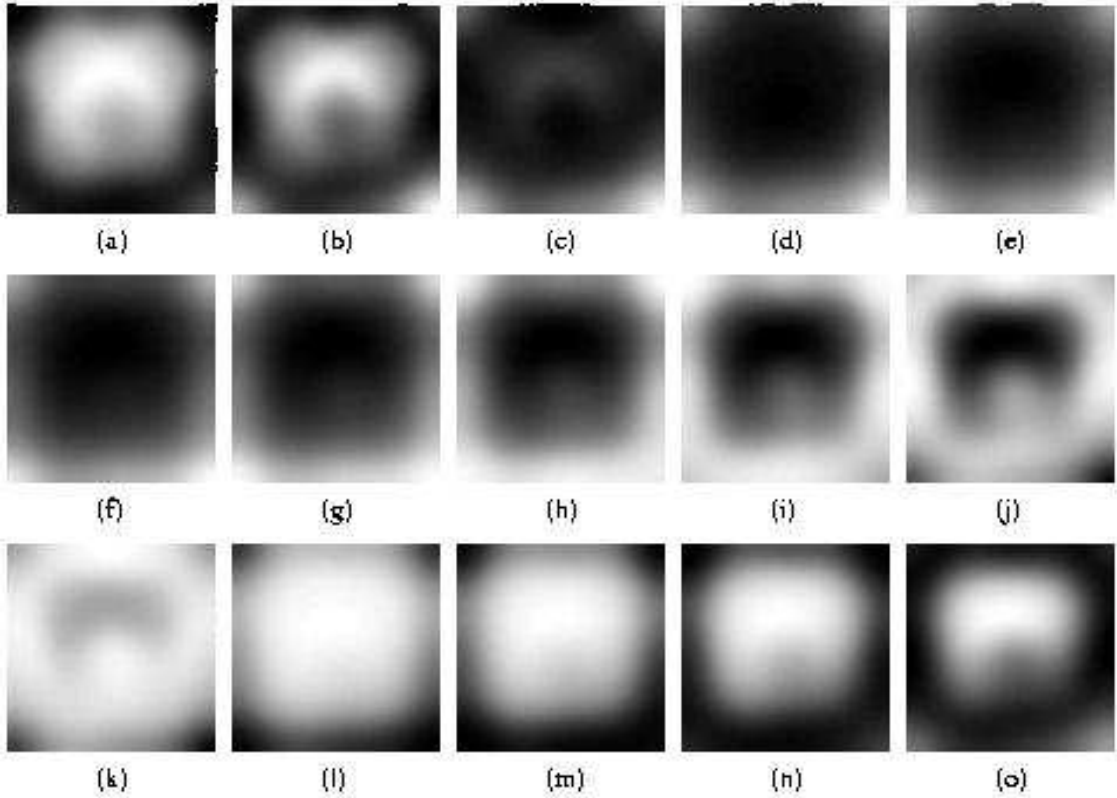}
\caption{
Snapshots showing breathing oscillations for $F=0.05$ and $\{ \beta,$ 
$\sigma,$ $\epsilon,$ $m_0$, $m_1,$ $m_2\}$ $=$ $\{0.734722,$ $0.734722,$ 
$0.15,$ $-0.874,$ $-0.4715,$ $-0.4715\}$ with $D_1=2$ and $D_2=5$ for various
time units starting from $T=4001$.
}
\label{fig9}
\end{figure}

In the above, we have shown that the inclusion of the external periodic
force can make a transition from one stationary pattern to another
stationary pattern like the transition from hexagons to rolls.  Besides
these, are there any time varying patterns? As mentioned above,
patterns such as localized and breathing oscillations have considerable
physical interest. In this regard, we considered the parameters $\{
\beta, \sigma, \epsilon, m_0, m_1, m_2\}$ $=$ $\{0.734722,$ $0.734722,$
$0.10,$ $-0.874,$ $ -0.4715,$ $-0.4715\}$ with $D_1=2.0$ and $D_2=5.0$
such that a regular hexagonal pattern is observed in the {\em absence}
of external periodic force.  From numerical simulations, we observed
that a space-time periodic oscillatory pattern (breathing motion) sets
in for a range of low values of $F$. Fig. \ref{fig9} shows the typical
snapshots of the oscillating pattern at various instants for the
specific choice $F=0.05$.  We have integrated over 10000 time units and
the figure corresponds to the region $T=4000 - 4014$. Typically we find
that the breathing pattern repeats itself approximately after a period
$T=15.0$ in the range of our integration. One may conclude that the
emergence of such breathing oscillations is due to the competition
between the Turing and Hopf modes in the presence of external periodic
force.

\section{Spatiotemporal chaos} 
\label{sec5}

Next we move on to a study of the spatiotemporal chaotic dynamics of
the array of coupled MLC circuits when individual cells are driven by
external periodic force. The motivation is that over a large domain of
($F$, $\omega$) values the individual MLC circuits typically exhibit
various bifurcations and transition to chaotic motion.  So one would
like to know how the coupled array behaves collectively in such a
situation, for fixed values of the parameters.  For this purpose, we
set the parameters at $\{\beta, \sigma, \epsilon, m_0, m_1, $ $m_2,
\omega\} = \{1.0, 1.015, 0, -1.02, -0.55, -0.55, 0.75 \}$. The
uncoupled systems exhibit period doubling bifurcations and chaotic
dynamics in the presence of external force. In our numerical
simulations, we have mainly considered the one dimensional array
specified by Eq. (1) and assumed periodic boundary conditions. 

\subsection{Spatiotemporal regular and chaotic motion}
\label{sec5:1}

\begin{figure}[!ht]
\centering
\includegraphics[width=\linewidth]{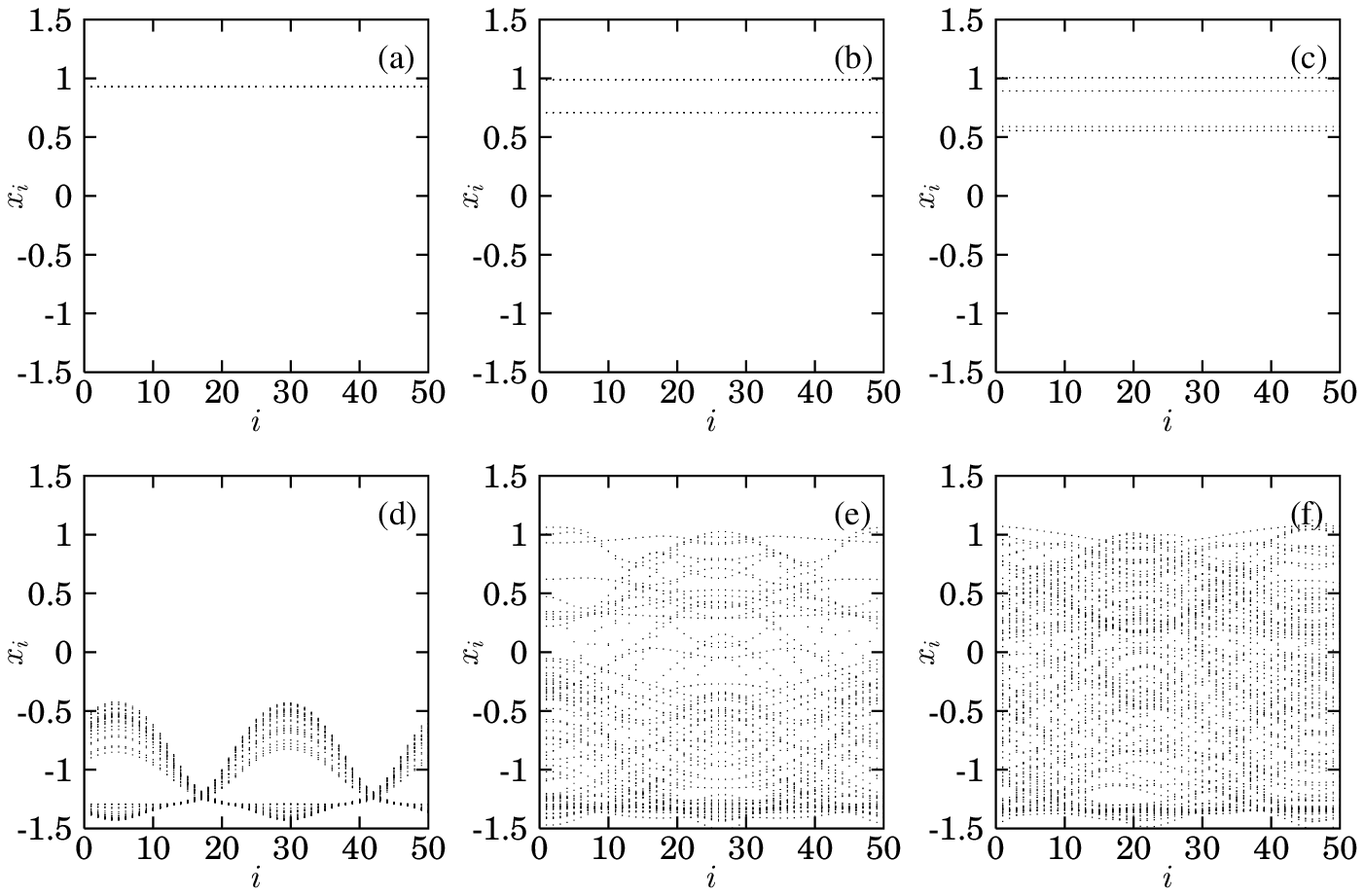}
\caption{
Space-amplitude plot showing the spatiotemporal periodic and chaotic
oscillations in 50 coupled MLC circuits for various values of external
periodic forcing strength: (a) $F=0.05$, (b) $F=0.08$, (c) $F=0.091$, 
(d) $F=0.12$, (e) $F=0.13$ and (f) $F=0.15$.
}
\label{fig10}
\end{figure}

Numerical simulations were performed by considering 50 cells and random
initial conditions using fourth order Runge-Kutta method for six
choices of $F$ values. The coupling coefficient in Eq. (\ref{eqn1all})
was chosen as $D=1.0$.  Out of these,  the first three lead to
period-$T$, period-$2T$, period-$4T$ oscillations, respectively and the
remaining choices correspond to chaotic dynamics of the single MLC
circuit.  Figs.~\ref{fig10}(a)-\ref{fig10}(g) show the space-time
plots for $F=0.05$, $F=0.08$, $F=0.09$, $F=0.12$, $F=0.13$ and
$F=0.15$, respectively. From the Figs. \ref{fig10}(a)-\ref{fig10}(c),
it can be observed that for $F=0.05$, $0.08$ and $0.09$ the MLC array
also exhibits regular periodic behaviour with periods $T$, $2T$ and
$4T$ respectively in time alone as in the case of the single MLC
circuit.  However, for $F=0.12$ and $0.13$(Figs. \ref{fig10}(d) and
\ref{fig10}(e)) one obtains {\em space-time periodic} oscillations
eventhough each of the individual uncoupled MLC circuits for the same
parameters exhibits chaotic dynamics. We may say that a kind of
controlling of chaos occurs due to the coupling, though the coupling
strength is large here.  From the above analysis it can be seen that
the macroscopic system shows regular behaviour in spite of the fact
that the microscopic subsystems oscillate chaotically.

Finally for $F=0.15$ the coupled system shows spatiotemporal chaotic
dynamics (Fig.  \ref{fig10}(f)) and this was confirmed by calculating
the Lyapunov exponents using the algorithm given by Wolf {\it et
al}\cite{wolf}. For example, we calculated the Lyapunov exponents for
$N=50$ coupled oscillators and we find the largest three exponents have
the values $\lambda_{\mbox{max}}=\lambda_0 =0.1001$,
$\lambda_1=0.0776$, $\lambda_2=0.0092$ and the rest are negative (see
next subsection for further analysis).

\subsection{Size instability, chaos synchronization and suppression of
STC}
\label{sec5:2}

Since the above study of spatiotemporal chaos involves a large number
of coupled chaotic oscillators, it is of great interest to analyse the
size dependence of the dynamics of these systems. To start with, we
consider the case of 10 coupled oscillators with periodic boundary
conditions and numerically solve the system with the other parameters
chosen as in Sec. \ref{sec5:1}. The value of $F$ is chosen in the range
$(0.12,0.15)$.  We find that this set up shows a different behaviour as
compared to the 50 cells case. Actually the system gets synchronized to
a chaotic orbit rather than showing periodic behaviour or
spatiotemporal chaos as in the case of 50 cells described above.

\begin{figure}[!ht]
\centering
\includegraphics[width=0.8\linewidth]{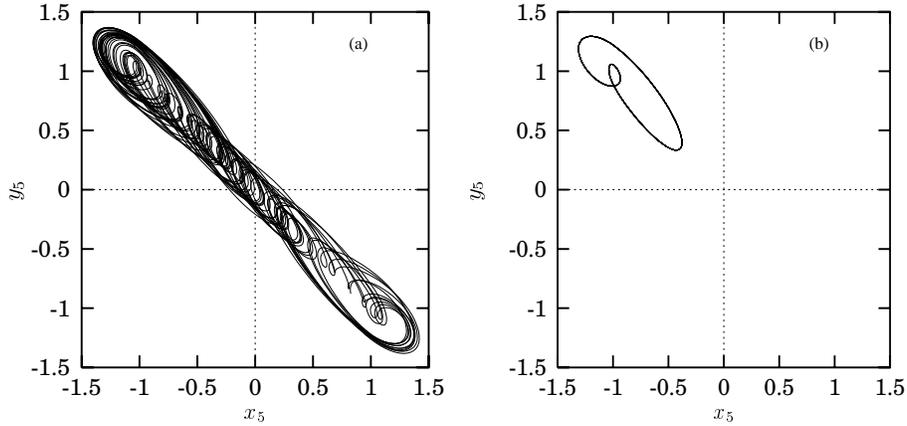}
\caption{
(a) The chaotic attractor at the $5^{\mbox{th}}$ cell for the synchronized
state $(N=42)$ and (b) the periodic orbit in the $5^{\mbox{th}}$ cell
for the controlled state $(N=43)$
}
\label{fig11}
\end{figure}

To start with we analyse the dynamics for $F=0.12$ by slowly increasing
the system size from $N=10$. It has been found that the coupled system
(\ref{eqn1all}) shows synchronized motion for $N\le 42$. This was
confirmed by calculating the Lyapunov spectrum which shows only one
positive exponent with the rest being negative. For example, for
$N=42$, $\lambda_{\mbox{max}}=0.1162$, with the rest of the exponents
being negative. The existence of only one positive exponent is a
necessary condition to have chaos synchronization\cite{peco}. Fig.
\ref{fig11}(a) shows the dynamics of the $5^{\mbox{th}}$ cell in the
array. The system shows entirely different behaviour when we increase
the system size to $N=43$.  As noted in the previous subsection,
\ref{sec5:1}, there occurs a kind of suppression of spatiotemporal
chaos.  Fig.~\ref{fig11}(b) shows the resultant periodic orbit in the
$5^{\mbox{th}}$ cell of the array.  The maximal Lyapunov exponent is
found to be negative in this case $(\lambda_{\mbox{max}} =
-0.001474)$.  Similar phenomenon has been observed for $F=0.13$ also.

Next we consider the case of $F=0.15$ in which the coupled system in
the previous subsection showed spatiotemporal chaos. From numerical
simulations, we again observed a synchronized motion for $N\le 31$ and
the corresponding Lyapunov spectrum shows one positive exponent only
with all the other exponents being negative. But for $N>31$, the coupled
system shows spatiotemporal chaos. The Lyapunov spectrum in this case
(for $N>31$) possesses multiple positive exponents. For example, for
$N=43$, $(\lambda_{\mbox{max}}=\lambda_0 =0.0997$, $\lambda_1=0.0633$,
$\lambda_2=0.0038)$.

The above type of size instability behaviour has also been found in a
coupled R\"ossler system as well which undergoes a short wavelength
bifurcation\cite{heagy}. In such cases, one can find the exact value of
the size below which stable synchronous oscillations occur.  However,
we find that the coupled MLC circuits do not show short wavelength
bifurcations. To arrive at a criteria for the size instability we
computed the Lyapunov dimension. It has been noted that the fractal
dimension per unit size, dimension density, is an appropriate quantity
for the spatiotemporal chaotic systems\cite{egolf,bohr}. A detailed
analysis on the size instability and chaos synchronization in coupled
system (\ref{eqn1all}) will be reported elsewhere.

\section{Conclusions}
\label{sec6}

In this lecture/article, we have tried to point out how reactive
diffusive nonliear systems can give rise to a wide variety of
spatiotemporal patterns ranging from trivial homogeneous states and
travelling wavefronts to Turing patterns and spatiotemporal chaos.
However even wider phenomena can be captured by discretized systems in
the form of cellular neural/nonlinear networks. As typical examples we
considered one and two dimensional arrays of coupled MLC circuits and
showed a wavefront propagation failure can occur due to certain
subcritical bifurcations. The onset of various interesting patterns
including Turing patterns such as hexagons, rhombs, rolls and the
effect of external forcing on them leading to breathing oscillations
were demonstrated. Finally how transitions to spatiotemporal chaos and
synchronization occurs and the role of system size in these transitions
have been brought out.

What has been demonstrated here is only a miniscule of the various
phenomena on a space-time scale which can occur in nonlinear reactive
diffusive systems in general and CNNs in particular. Even these basic
patterns can play a very useful role in developing synchronized
communication systems, patterns recognition, image processing and so
on.  Much more work need to be done in order discern all basic patterns
which can arise in CNNs and to understand their structure and
stability. This can in turn will lead to a better understanding of
nonlinear dynmaical systems in general.

\section*{Acknowledgements}

The work of M.L. forms part of Department of Science \& Technology
sponsored Govt. of India research project. The work of P.M. has been
supported by Council of Scientific \& Industrial Research through a
Senior Research Fellowship.

\end{document}